\documentclass{moriond}
\usepackage{hyperref}

\newcommand*{\doi}[2]{\href{http://dx.doi.org/#1}{#2}}
\def\Journal#1#2#3#4{{#1} {\bf #2}, #3 (#4)}

\def\NPB{{\em Nucl. Phys.} B}

\def\PRL{\em Phys. Rev. Lett.}
\def\PRD{{\em Phys. Rev.} D}

\def\be{\begin{equation}}
\def\ee{\end{equation}}
\def\bea{\begin{eqnarray}}
\def\eea{\end{eqnarray}}

\newcommand{\Photo}{}

\begin{document}
\vspace*{4cm}
\title{SOME COSMOLOGICAL CONSEQUENCES OF A BREAKING OF THE EINSTEIN EQUIVALENCE PRINCIPLE}

\author{ A. HEES$^{1}$, O. MINAZZOLI$^2$, J. LARENA$^1$ }

\address{$^1$ Department of Mathematics, Rhodes University, 6139 Grahamstown, South Africa\\
$^2$ Centre Scientifique de Monaco -- 
 UMR ARTEMIS, CNRS, University of Nice Sophia-Antipolis, France}

\maketitle\abstracts{In this communication, we consider a wide class of extensions to General Relativity that break explicitly the Einstein Equivalence Principle by introducing a multiplicative coupling between a scalar field and the electromagnetic Lagrangian. In these theories, we show that 4 cosmological observables are intimately related to each other: a temporal variation of the fine structure constant, a violation of the distance-duality relation, the evolution of the cosmic microwave background (CMB) temperature and CMB spectral distortions. This enables one to put very stringent constraints on possible violations of the distance-duality relation, on the evolution of the CMB temperature and on admissible CMB spectral distortions using current constraints on the fine structure constant. Alternatively, this offers interesting possibilities to test a wide range of theories of gravity by analyzing several datasets concurrently.}

\section{Introduction}
The theory of General Relativity (GR) is based upon two fundamental principles: the Einstein Equivalence Principle (EEP) which gives to gravitation its geometrical nature and the Einstein field equations that specify the form of the space-time geometry. All GR extensions (in 4 dimensions) produce a deviation from at least one of these principles and it is therefore highly important to test and to constrain them (see~\cite{will:2014la} and references therein for a review of the tests of GR). From a theoretical point of view, the EEP implies the existence of a space-time metric $g_{\mu\nu}$ to which the matter Lagrangian is minimally coupled to~\cite{will:2014la}. Phenomenologically, three aspects of the EEP can be tested~\cite{will:2014la}: (i) the Universality of Free Fall, (ii) the Local Lorentz Invariance and (iii) the Local Position Invariance.

A way to break the EEP in tensor-scalar theory is to introduce a multiplicative coupling between the scalar field and the matter Lagrangian. This kind of coupling appears naturally in string-inspired theories~\cite{damour:1994fk,damour:2010zr}, in theories with a varying fine structure constant~\cite{bekenstein:1982zr,sandvik:2002ly} or in the so-called \emph{pressuron} phenomenology~\cite{minazzoli:2013fk,minazzoli:2014xz}.

In this communication, we will focus on the cosmological implications of such coupling in the electromagnetic sector
\begin{equation}\label{eq:action}
	S_\textrm{EM}=-\frac{1}{4}\int d^4x \sqrt{-g} e^{-2\varphi} F^{\mu\nu}F_{\mu\nu} + q_p\int A_\mu dx_p^\mu\, ,
\end{equation}
where $g$ is the determinant of the space-time metric $g_{\mu\nu}$, $\varphi$ is a scalar field whose action is unspecified, $F_{\mu\nu}=\partial_\mu A_\nu-\partial_\nu A_\mu$ is the standard Faraday tensor, $A_\mu$ is the 4-potential and $q_p$ is the electric charge of a particle interacting with the electromagnetic field. It is important to point out that in order to preserve the $U(1)$ gauge invariance, the interaction part of the electromagnetic Lagrangian can not include a contribution from the scalar field~\cite{bekenstein:1982zr,hees:2015yq}. The coupling introduced breaks explicitly the EEP (in particular, it can not be absorbed through a conformal transformation). Implications of this kind of couplings on the universality of free fall and on variations of fundamental constants have been studied for instance in \cite{damour:1994fk,damour:2010zr,damour:2012zr}. In addition to these effects, we will show that four cosmological observables are modified (with respect to GR) and are intimately related to each other in this class of theory~\cite{hees:2014uq}: (i) temporal variation of the fine structure constant, (ii) violation of the distance-duality relation, (iii) modification of the evolution of the CMB temperature and (iv) CMB spectral distortions.

\section{Modification of cosmological observables}
The coupling introduced in Eq.~\ref{eq:action} leads to a modification of four cosmological observables. Details concerning the theoretical derivations of these observables can be found in~\cite{hees:2014uq}. It is worth to insist on the fact that the derivation relies only on the matter part of the action and not on the gravitational part. This means that our results apply to a very wide class of gravitation theories. In a Friedman-Lema\^itre-Robertson-Walker space-time, the expressions of the four observables are given by the following expressions:
\begin{enumerate}
	\item temporal variation of the fine structure constant. A straightforward identification in the action leads to~\cite{uzan:2011vn} $\Delta\alpha/\alpha=(\alpha(z)-\alpha_0)/\alpha_0=e^{2(\varphi-\varphi_0)}-1\approx 2(\varphi-\varphi_0)$, where $z$ is the redshift and subscripts $0$ refer to $z=0$.
	\item violation of the cosmic distance-duality relation. The optic geometric limit of the modified Maxwell equations shows that photons propagate on null geodesics but their number is not conserved due to an exchange with the scalar field. Therefore, the expression of the angular diameter distance ($D_A$) is the same as in GR but this leads to a modification of the distance-luminosity expression ($D_L$)~\cite{minazzoli:2014xz} and hence to a violation of the cosmic distance-duality relation: $\eta(z)=D_L(z)/(D_A(z)(1+z)^2)=e^{(\varphi-\varphi_0)}\approx 1 +(\varphi-\varphi_0)$.
	\item modification of the evolution of the CMB temperature. Considering the CMB as a gaz of photons described by a distribution function solution of a relativistic Boltzman equation and using the geometric optic approximation of the modified Maxwell equation lead to a modification of the CMB temperature evolution: $$
	T(z)=T_0(1+z)\left[1+0.12\left(e^{2(\varphi-\varphi_{CMB})}-e^{2(\varphi_{0}-\varphi_{CMB})}\right)\right]\approx T_0(1+z)\left[1+0.24(\varphi-\varphi_0)\right].$$
	\item spectral distortion of the CMB. Using the same approach as the one sketched in the last item, one gets an expression for the chemical potential of the CMB radiation at current epoch $\mu=0.47\left(e^{2(\varphi_{CMB}-\varphi_{0})}-1\right)\approx 0.94 \left(\varphi_{CMB}-\varphi_0\right)$. 
\end{enumerate}
To summarize, the coupling introduced in Eq.~\ref{eq:action} implies that the four observables are intimately linked to each other through the relations
\begin{eqnarray}
	\varphi(z)-\varphi_0 &=& \frac{1}{2}\frac{\Delta\alpha(z)}{\alpha}=\eta(z)-1=4.17\left(\frac{T(z)}{T_0(1+z)}-1\right) \label{eq:phi}\\
	\mu &=& 0.47 \ \frac{\Delta\alpha(z_{CMB})}{\alpha}=0.94\left(\eta(z_{CMB})-1\right)=3.92\left(\frac{T(z_{CMB})}{T_0(1+z_{CMB})}-1\right). \label{eq:mu}
\end{eqnarray}

\section{Transformation of experimental constraints assuming a multiplicative coupling}
Assuming that the theory of gravitation is described by the multiplicative coupling introduced in Eq.~\ref{eq:action} (which is a large class of theory including GR), we can use Eq.~\ref{eq:phi} to transform observational constraints on one type of observations into constraints on another type. In this communication, we use three datasets of $\alpha$ measurements: precise clocks measurements of variations of $\alpha$ providing the constraint~\cite{rosenband:2008fk,guena:2012ys} $\dot\alpha/\alpha=(-1.6\pm 2.3)\times 10^{-17}\textrm{yr}^{-1}$, 154 quasar absorption lines observed at the VLT~\cite{king:2012hb} and 128 quasar absorption lines observed at the Keck observatory~\cite{murphy:2003dz}. Using separately these three datasets and Eq.~\ref{eq:phi}, we constrain the parameters $\eta_{i}$, $\varepsilon$ and $\beta$ that enter standard parametrizations of $\eta(z)$ and $T(z)$:
\begin{eqnarray*}
&&	\eta(z)=1+\eta_0,\phantom{z}   \qquad  \eta(z)=1+\eta_2\frac{z}{1+z},\phantom{(\ln)} \qquad \eta(z)=(1+z)^\varepsilon, \\
&&	\eta(z)=1+\eta_1z, \qquad \eta(z)=1+\eta_3\ln (1+z), \qquad T(z)=(1+z)^{1-\beta} .
\end{eqnarray*}
A Bayesian inversion of the three datasets lead to estimations of the $\eta_i$, $\varepsilon$ and $\beta$ parameters given in Tab.~\ref{tab}. The constraints derived from clocks measurements rely only on one observation and is valid only if no screening mechanism occurs around Earth. The obtained constraints improve by 5 orders of magnitude direct observations of $\eta$ or of the CMB temperature~\cite{planck-collaboration:2015fk} but are valid only under the assumption that the coupling given by Eq.~\ref{eq:action} is valid. 
\begin{table}[htb]
\caption[]{Estimation of the parameters entering standard parametrizations of $\eta(z)$ and $T(z)$ using Eq.~\ref{eq:phi} and measurements of $\alpha$ from three different datasets.}
\label{tab} 
\vspace{0.4cm}
\centering
\begin{tabular}{c c c c}
\hline
 Parameter  & \multicolumn{3}{c}{Estimation $[\times 10^{-7}]$} \\
 & VLT & Keck  & Clocks \\
\hline
$\eta_0$     &  $10\pm 6\phantom{0.}$ &  $-29\pm10$             &-\\
$\eta_1$       &  $8.4\pm 3.5$           &  $-16\pm6\phantom{1}$  & $1.0\pm1.4$\\
$\eta_2$       &  $20\pm 10\phantom{.}$  &  $-49\pm17$            & $1.0\pm1.4$\\
$\eta_3$       &  $14\pm 6\phantom{0.}$  &  $-30\pm11$            & $1.0\pm1.4$\\
$\varepsilon$  & $14\pm 6\phantom{0.}$   &  $-30\pm 11$           & $1.0\pm1.4$ \\\hline
$\beta$  & $-3.3\pm 1.5$   &  $7.2\pm 2.5$           & $-0.3\pm 0.3$ \\
\hline
\end{tabular}
\end{table}

\section{Test of the multiplicative coupling using Gaussian processes}
We can also use the different types of observations to assess the validity of the coupling introduced in Eq.~\ref{eq:action}. Indeed, if different types of observations indicate a violation of Eqs.~\ref{eq:phi}-\ref{eq:mu}, this would be an indication of a departure from a multiplicative coupling. Here, we transform separately observations on $\Delta\alpha/\alpha$, on $\eta(z)$ and on the CMB temperature into an estimation of the evolution of $\varphi-\varphi_0$ using Eq.~\ref{eq:phi}. Then, we compare the different evolutions of the scalar field estimated from the different types of observations to see if they are compatible. The analysis is done using Gaussian processes~\cite{seikel:2012fk} in order to provide a model independent analysis. More details on the analysis can be found in~\cite{hees:2014uq}. 
\begin{figure}[hb]
\includegraphics[width=0.45\linewidth]{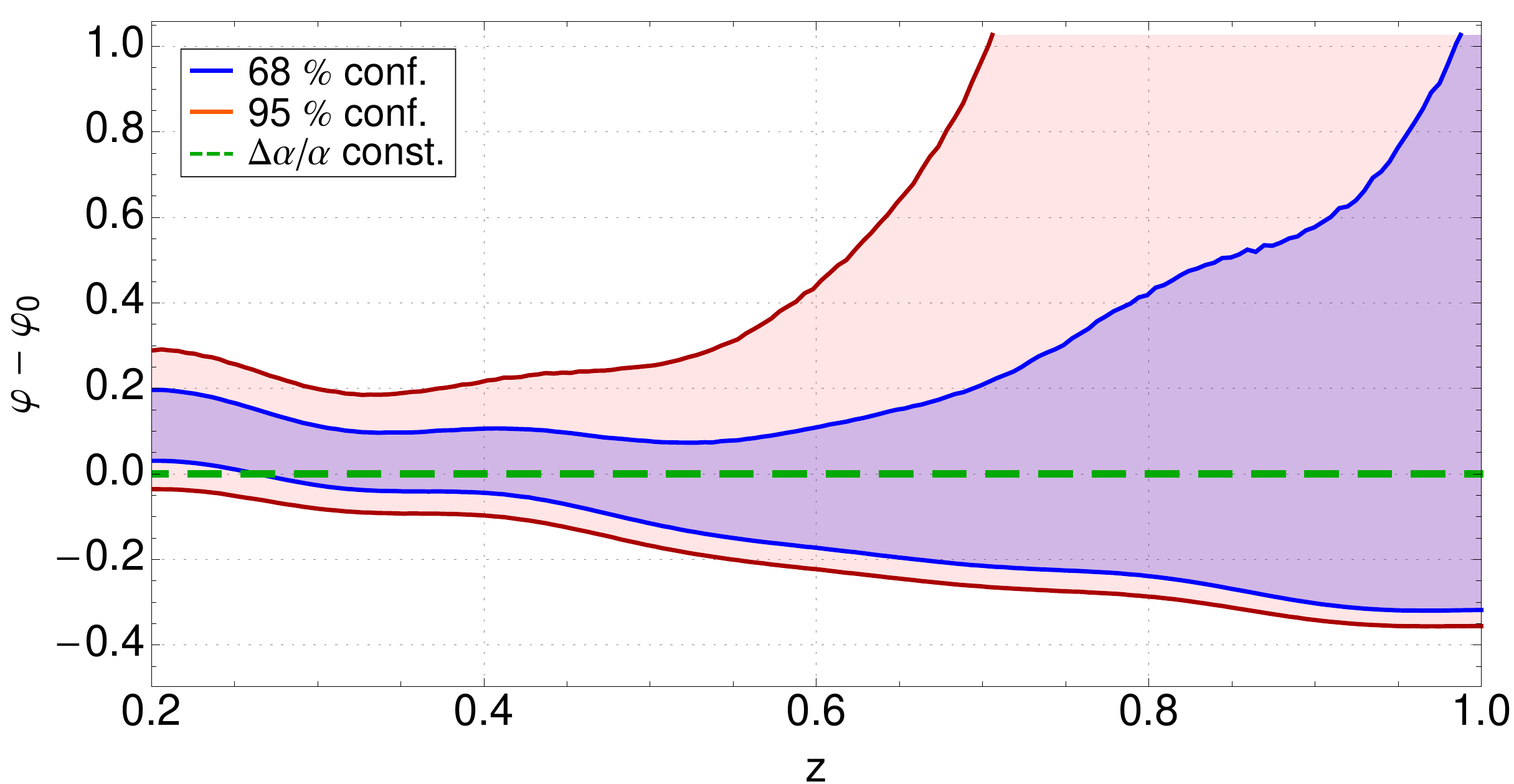}
\hfill
\includegraphics[width=0.45\linewidth]{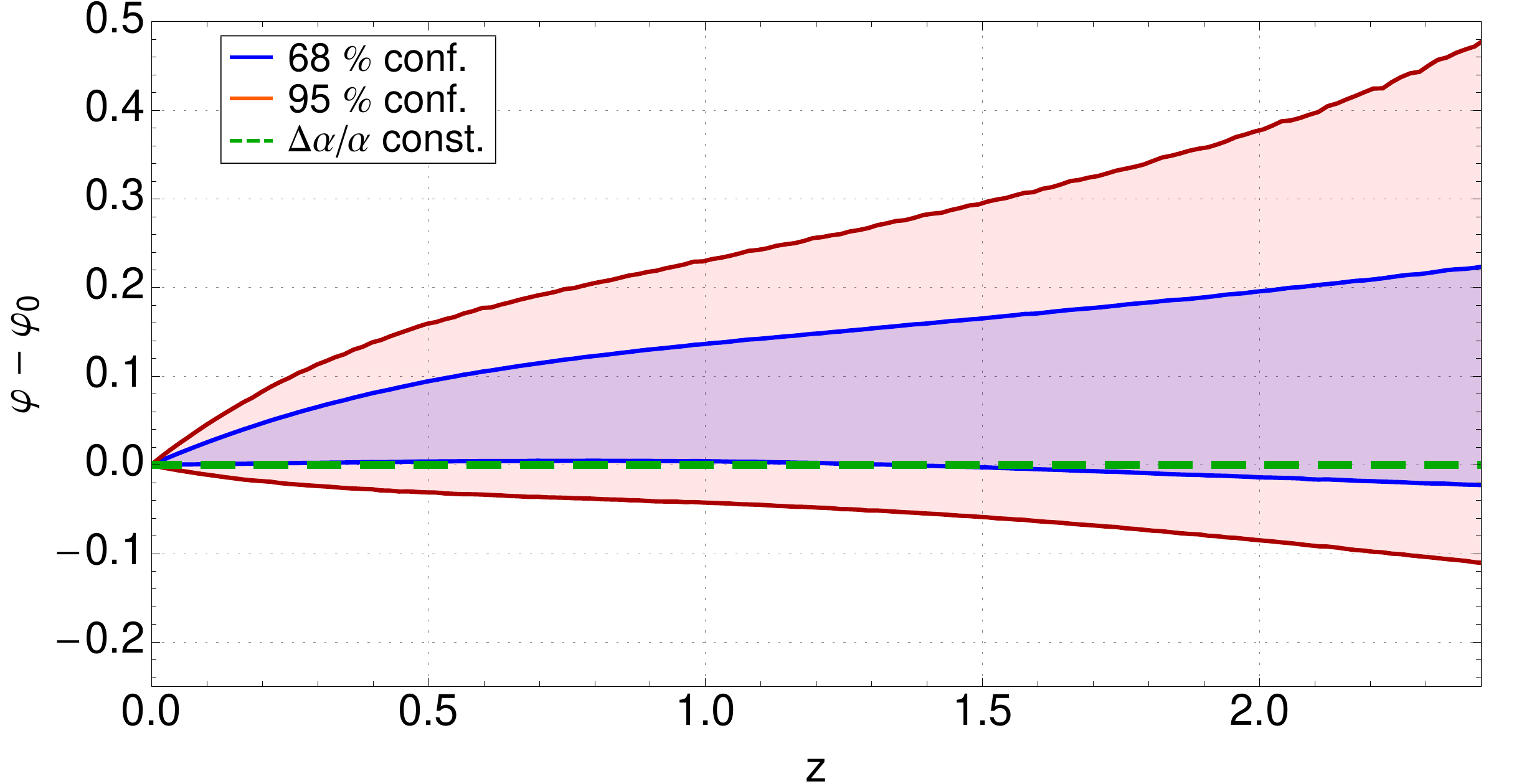}
\caption[]{Estimation of the evolution of $\varphi-\varphi_0$ from observations of $\Delta\alpha/\alpha$ (green dashed lines), from observations of angular and luminosity distance (left) and from observations of CMB temperature (right) using Eq.~\ref{eq:phi}.}
\label{fig:gp}
\end{figure}

On Fig.~\ref{fig:gp} is represented the  evolution of the scalar field estimated from observations of $\Delta\alpha/\alpha$ (dashed green lines), from observations of angular and luminosity distance~\cite{bonamente:2006lq,de-filippis:2005pd} (on the left of the figure) and from observations of the CMB temperature~\cite{saro:2014ai} (on the right of the figure). The comparison of the three different evolutions does not show any incoherence. Therefore, current data are consistent with the coupling considered in Eq.~\ref{eq:action}. One  limitation for this test comes from angular distance measurements. We estimate that EUCLID and the SKA will improve this test by one order of magnitude by measuring the BAO at different redshifts~\cite{hees:2014uq}.

\section{Conclusion}
A multiplicative coupling between a scalar field and the electromagnetic Lagrangian produces a violation of the EEP. Amongst others, it is known that this type of coupling leads to a violation of the universality of the free fall~\cite{damour:1994fk,damour:2010zr,damour:2012zr} and to variations of the ``constants'' of Nature~\cite{uzan:2011vn,damour:2012zr}. In this communication, we show that this type of coupling has also some cosmological implications. In particular, it will produce four cosmological deviations with respect to GR at the cosmological scales: temporal variation of $\alpha$, violation of the distance-duality relation, modification of the evolution of the CMB temperature and CMB spectral distortions. Therefore, in this class of models, these cosmological observations are complementary to local constraints on the EEP.

In addition, we have shown that in this class of theory, there are unambiguous relations between these four observables. These relations allow one to transform measurements of $\Delta\alpha/\alpha$ on constraints on $\eta(z)$ and on the evolution of the CMB temperature. Using current data, this results in an improvement by 5 orders of magnitude in the estimation of the parameters entering standard expressions of $\eta(z)$ and CMB temperature evolution. This improvement is only valid under the assumption that the coupling introduced in Eq.~\ref{eq:action} holds. 

Finally, a comparison between the different types of observations leads to a test of the multiplicative coupling introduced in Eq.~\ref{eq:action}. Current data are compatible with this coupling.

\section*{Acknowledgments}

AH and OM thank the organizers for financial support to attend the conference. In addition, AH acknowledges support from ``Fonds Sp\'ecial de Recherche" through a FSR-UCL grant. JL is supported by the National Research Foundation (South Africa).


\end{document}